\documentclass[preprint2]{emulateapj}

\usepackage{natbib}
\usepackage{longtable}
\usepackage[]{graphicx}
\usepackage{amsmath}
\usepackage{tabularx}
\usepackage{bm}
\usepackage{color}
\usepackage{hyperref}

\shorttitle{The $M_{\rm BH} - n_{\rm sph}$ correlation}
\shortauthors{Savorgnan}

\begin{document}

\title{Supermassive black holes and their host spheroids \\ 
III. The $M_{\rm BH} - \lowercase{n_{\rm sph}}$ correlation }

\author{Giulia A.~D.~Savorgnan\altaffilmark{1}}
\affil{Centre for Astrophysics and Supercomputing, Swinburne University of Technology, Hawthorn, Victoria 3122, Australia.}
\email{gsavorgn@astro.swin.edu.au}

\begin{abstract}
The S\'ersic $R^{1/n}$ model is the best approximation known to date for describing the light distribution 
of stellar spheroidal and disk components, 
with the S\'ersic index $n$ providing a direct measure of the central radial concentration of stars. 
The S\'ersic index of a galaxy's spheroidal component, $n_{sph}$, has been shown to tightly correlate 
with the mass of the central supermassive black hole, $M_{BH}$. 
The $M_{BH}-n_{sph}$ correlation is also expected from other two well known scaling relations 
involving the spheroid luminosity, $L_{sph}$: 
the $L_{sph}-n_{sph}$ and the $M_{BH}-L_{sph}$. 
Obtaining an accurate estimate of the spheroid S\'ersic index requires a careful modelling of a galaxy's light distribution 
and some studies have failed to recover a statistically significant $M_{BH}-n_{sph}$ correlation. 
With the aim of re-investigating the $M_{BH}-n_{sph}$ 
and other black hole mass scaling relations, 
we performed a detailed (i.e.~bulge, disks, bars, spiral arms, rings, halo, nucleus, etc.) 
decomposition of 66 galaxies, with directly measured black hole masses, 
that had been imaged at $3.6\rm~\mu m$ with \emph{Spitzer}. 
In this paper, the third of this series, we present an analysis of the $L_{sph}-n_{sph}$ 
and $M_{BH}-n_{sph}$ diagrams. 
While early-type (elliptical+lenticular) and late-type (spiral) galaxies split into two separate relations 
in the $L_{sph}-n_{sph}$ and $M_{BH}-L_{sph}$ diagrams, 
they reunite into a single $M_{BH} \propto n_{sph}^{3.39 \pm 0.15}$ sequence
with relatively small intrinsic scatter ($\epsilon \simeq 0.25 \rm~dex$). 
The black hole mass appears to be closely related to the spheroid central concentration of stars, 
which mirrors the inner gradient of the spheroid gravitational potential. 
\end{abstract}


\section{Introduction}
The empirical \cite{sersic1963,sersic1968} $R^{1/n}$ model has been demonstrated to provide adequate description 
of the light distribution of the stellar spheroidal\footnote{Throughout the text, we use the term ``spheroid'' to indicate 
either a disk-less elliptical galaxy or the bulge component of a disk galaxy;  
we do not attempt at distinguishing between classical bulges and disk-like pseudo-bulges. } and disk components of galaxies 
(e.g.~\citealt{caon1993,andredakis1995,iodice1997,iodice1999,seigar1998,khosroshahi2000}), 
yet its physical origin has remained unexplained for decades. 
The S\'ersic model parameterizes the intensity of light $I$ as a function of the projected galactic radius $R$ such that
\begin{equation*}
I(R; I_{\rm e},R_{\rm e},n) = I_{\rm e} \exp \biggl \{ -b_{\rm n} \biggl [ \biggl (\frac{R}{R_{\rm e}} \biggr )^{1/n} -1 \biggr ] \biggr \}, 
\end{equation*}
where $I_{\rm e}$ indicates the intensity at the effective radius $R_{\rm e}$ that encloses half of the total light from the model, 
the S\'ersic index $n$ is the parameter that regulates the curvature of the radial light profile, 
and $b_{\rm n}$ is a constant defined in terms of the S\'ersic index (see \citealt{grahamdriver2005}, and references therein). 
A large S\'ersic index corresponds to a steep inner profile and a shallow outer profile, 
whereas a small S\'ersic index corresponds to a shallow inner profile and a steep outer profile. 
This means that, for a stellar spheroidal system whose light distribution is well approximated by the S\'ersic model, 
the larger the S\'ersic index is, the more centrally concentrated the stars are and the more extended the outer envelope is. \\
A compelling physical interpretation for the S\'ersic profile family was recently theorized by \cite{cen2014} 
and later confirmed by \cite{nipoti2015} by means of $N$-body simulations. 
\cite{cen2014} conjectured that, when structures form within a standard cold dark matter model seeded by random Gaussian fluctuations, 
any centrally concentrated stellar structure always possesses an extended stellar envelope, and vice versa. 
\cite{nipoti2015} quantitatively explored Cen's hypothesis and showed that 
systems originated from several mergers have a large S\'ersic index ($n \gtrsim 4$), 
whereas systems with a S\'ersic index as small as $n \simeq 2$ can be produced by coherent dissipationless collapse, 
and exponential profiles ($n=1$) can only be obtained through dissipative processes.
This scenario sets the theoretical framework for the well known correlation between the spheroid luminosity, $L_{\rm sph}$,  
and the spheroid S\'ersic index, $n_{\rm sph}$, (e.g.~\citealt{youngcurrie1994,jerjen2000,grahamguzman2003}), 
although the numerical results of \cite{nipoti2015} seem to lack of spheroidal systems with S\'ersic indices as large as $7 - 10$, 
which are commonly observed in the local Universe.  \\
Given the existence of the $L_{\rm sph} - n_{\rm sph}$ correlation 
and the relation between the central black hole mass, $M_{\rm BH}$, and the spheroid luminosity 
(e.g.~\citealt{kormendyrichstone1995,magorrian1998,marconihunt2003,haringrix2004}), 
an $M_{\rm BH} - n_{\rm sph}$ relation must exist. 
After \cite{graham2001bulges} showed that the black hole mass is tightly linked to the stellar light concentration of spheroids 
(measured through a parameter different from, but closely related to the S\'ersic index), 
\cite{grahamdriver2007} presented for the first time the $M_{\rm BH} - n_{\rm sph}$ correlation 
using a sample of 27 elliptical and disk galaxies. 
\cite{grahamdriver2007} fit their data with a log-quadratic regression, 
finding that the $M_{\rm BH} - n_{\rm sph}$ log-relation is steeper for spheroids with small S\'ersic indices 
and shallower for spheroids with large S\'ersic indices, 
and measured a relatively small level of scatter\footnote{At the time, the $M_{\rm BH} - \sigma$ relation 
\citep{ferraresemerritt2000,gebhardt2000} was reported to have the same level of scatter as the $M_{\rm BH} - n_{\rm sph}$ 
relation ($\simeq 0.3~\rm dex$). }.
A few years later, \cite{sani2011}, \cite{vika2012} and \cite{beifiori2012} performed multi-component decompositions 
for samples of galaxies similar to that used by \cite{grahamdriver2007}, 
but they failed to recover a strong $M_{\rm BH} - n_{\rm sph}$ relation. 
This issue was tackled by \cite{savorgnan2013}, who collected the S\'ersic index measurements published by 
\cite{grahamdriver2007}, \cite{sani2011}, \cite{vika2012} and \cite{beifiori2012} for a sample of 54 galaxies, 
and showed that, by rejecting the most discrepant measurements and averaging the remaining ones, 
a strong $M_{\rm BH} - n_{\rm sph}$ relation was retrieved. 
Remarkably, \cite{savorgnan2013} repeated their analysis upon excluding the S\'ersic index measurements of \cite{grahamdriver2007} 
and still regained a significant $M_{\rm BH} - n_{\rm sph}$ correlation. 
This was suggesting that the individual galaxy decompositions of \cite{sani2011}, \cite{vika2012} and \cite{beifiori2012} were not accurate, 
i.e.~each individual study obtained ``noisy'' S\'ersic index measurements 
which prevented the recovery of a strong $M_{\rm BH} - n_{\rm sph}$ relation. \\
Motivated by the need for more accurate galaxy decompositions to refine and re-investigate scaling relations 
between the black hole mass and several host spheroid structural parameters, 
we performed state-of-the-art modelling for the largest sample of galaxies to date (\citealt{paperI}, hereafter \emph{Paper I})  
for which a dynamical measurement of the black hole mass was available.
In doing so, we used $3.6\rm~\mu m$ \emph{Spitzer} satellite imagery, 
given its superb capability to trace the stellar mass (\citealt{sheth2010}, and references therein). 
In \citeauthor{paperII} (\citeyear{paperII}, hereafter \emph{Paper II}) we examined the correlations between the black hole mass and 
the total galaxy luminosity, the spheroid luminosity and the spheroid stellar mass. 
Here we focus on the $M_{\rm BH} - n_{\rm sph}$ relation.

\section{Data}
We populated the $L_{\rm sph} - n_{\rm sph}$ and $M_{\rm BH} - n_{\rm sph}$ diagrams 
with the same galaxy sample used in \emph{Paper II} (and presented here in Table \ref{tab:sample}), 
i.e.~66 galaxies for which a dynamical measurement of the black hole mass has been reported in the literature 
(by \citealt{grahamscott2013} or \citealt{rusli2013}) 
and for which we were able to successfully model the light distribution and measure the spheroid structural parameters 
using $3.6\rm~\mu m$ \emph{Spitzer} satellite images. 
Our galaxy decompositions take into account bulge, disks, spiral arms, bars, rings, halo, 
extended or unresolved nuclear source and partially depleted core, 
and -- for the first time -- they were checked to be consistent with the galaxy kinematics 
\citep{atlas3dIII,scott2014,arnold2014}. 
Kinematical information was used to confirm the presence of disk components 
in the majority of early-type (elliptical + lenticular) galaxies and, more importantly,  
to establish the radial extent of these disks, 
which in most cases is not obvious from a visual inspection of the galaxy images. 
This enabled us to distinguish between intermediate-scale disks, 
that are fully embedded in the spheroid,  
and large-scale disks, that encase the bulge and dominate the light at large radii.  
\cite{ellicular} demonstrate that when an intermediate-scale disk is misclassified and modeled as a large-scale disk, 
the luminosity of the spheroid is underestimated, 
hence the galaxy incorrectly appears as a positive outlier (an ``over-massive'' black hole) in the $M_{\rm BH} - L_{\rm sph}$ diagram. 
A detailed description of the dataset used here, the data reduction process and the galaxy modelling technique that we developed 
can be found in \emph{Paper I}, 
along with a discussion of how we estimated the uncertainties on the spheroid S\'ersic indices\footnote{The uncertainties associated with 
the spheroid S\'ersic indices were estimated with a method that takes into account systematic errors. 
This method consists in comparing, for each of our galaxies, the measurements of the spheroid S\'ersic index obtained by other studies 
with that obtained by us. 
Systematic errors are typically not considered by popular 2D fitting codes, which report only the statistical errors 
associated with their fitted parameters. 
Readers should refer to \emph{Paper I} for a more detailed discussion on this topic. }. 
For the present analysis, we elected to use the spheroid S\'ersic indices obtained from the decomposition of the one-dimensional 
major-axis surface brightness profiles of our galaxies.  
The morphological classification (E = elliptical; E/S0 = elliptical/lenticular; S0 = lenticular; 
S0/Sp = lenticular/spiral; Sp = spiral; and ``merger'') follows from the galaxy decompositions illustrated in \emph{Paper I}. 
As in \emph{Paper II}, we will refer to early-type galaxies (E+S0) and late-type galaxies (Sp). 
The early-type bin includes the two galaxies classified as E/S0, 
whereas the two galaxies classified as S0/Sp and the two galaxies classified as ``mergers'' are included in neither the early- nor the late-type bin.

\begin{table*}                                        
\small                                                
\begin{center}                                        
\caption{Galaxy sample.} 
\begin{tabular}{llllllrll}                           
\tableline                                                
\multicolumn{1}{l}{{\bf Galaxy}} &                   
\multicolumn{1}{l}{{\bf Type}} &                     
\multicolumn{1}{l}{{\bf Distance}} &                 
\multicolumn{1}{l}{{\bf $\bm{M_{\rm BH}}$}} &  
\multicolumn{1}{l}{{\bf $\bm{MAG_{\rm sph}}$}} &  
\multicolumn{1}{l}{{\bf $\bm{n_{\rm sph}^{\rm maj}}$}} \\  
\multicolumn{1}{l}{} &                                
\multicolumn{1}{l}{} &                                
\multicolumn{1}{l}{[Mpc]} &                           
\multicolumn{1}{l}{$[10^8~\rm M_{\odot}]$} &         
\multicolumn{1}{l}{[mag]} &                                
\multicolumn{1}{l}{} \\                             
\multicolumn{1}{l}{(1)} &                             
\multicolumn{1}{l}{(2)} &                             
\multicolumn{1}{l}{(3)} &                             
\multicolumn{1}{l}{(4)} &                             
\multicolumn{1}{l}{(5)} &                             
\multicolumn{1}{l}{(6)} \\  
\tableline                                                
IC 1459  &  E  &  $28.4$  &  $24_{-10}^{+10}$   &  $-26.15_{-0.11}^{+0.18}$   &  $6.6_{-0.8}^{+0.9}$   &   \\ 
IC 2560  &  Sp (bar)  &  $40.7$  &  $0.044_{-0.022}^{+0.044}$   &  $-22.27_{-0.58}^{+0.66}$   &  $0.8_{-0.3}^{+0.4}$   &   \\ 
IC 4296  &  E  &  $40.7$  &  $11_{-2}^{+2}$   &  $-26.35_{-0.11}^{+0.18}$   &  $5.8_{-0.7}^{+0.8}$   &   \\ 
M31  &  Sp (bar)  &  $0.7$  &  $1.4_{-0.3}^{+0.9}$   &  $-22.74_{-0.11}^{+0.18}$   &  $2.2_{-0.3}^{+0.3}$   &   \\ 
M49  &  E  &  $17.1$  &  $25_{-1}^{+3}$   &  $-26.54_{-0.11}^{+0.18}$   &  $6.6_{-0.8}^{+0.9}$   &   \\ 
M59  &  E  &  $17.8$  &  $3.9_{-0.4}^{+0.4}$   &  $-25.18_{-0.11}^{+0.18}$   &  $5.5_{-0.7}^{+0.8}$   &   \\ 
M64  &  Sp  &  $7.3$  &  $0.016_{-0.004}^{+0.004}$   &  $-21.54_{-0.11}^{+0.18}$   &  $0.8_{-0.1}^{+0.1}$   &   \\ 
M81  &  Sp (bar)  &  $3.8$  &  $0.74_{-0.11}^{+0.21}$   &  $-23.01_{-0.66}^{+0.88}$   &  $1.7_{-0.7}^{+1.3}$   &   \\ 
M84  &  E  &  $17.9$  &  $9.0_{-0.8}^{+0.9}$   &  $-26.01_{-0.58}^{+0.66}$   &  $7.8_{-2.5}^{+3.6}$   &   \\ 
M87  &  E  &  $15.6$  &  $58.0_{-3.5}^{+3.5}$   &  $-26.00_{-0.58}^{+0.66}$   &  $10.0_{-3.2}^{+4.7}$   &   \\ 
M89  &  E  &  $14.9$  &  $4.7_{-0.5}^{+0.5}$   &  $-24.48_{-0.58}^{+0.66}$   &  $4.6_{-1.5}^{+2.2}$   &   \\ 
M94  &  Sp (bar)  &  $4.4$  &  $0.060_{-0.014}^{+0.014}$   &  $-22.08_{-0.11}^{+0.18}$   &  $0.9_{-0.1}^{+0.1}$   &   \\ 
M96  &  Sp (bar)  &  $10.1$  &  $0.073_{-0.015}^{+0.015}$   &  $-22.15_{-0.11}^{+0.18}$   &  $1.5_{-0.2}^{+0.2}$   &   \\ 
M104  &  S0/Sp  &  $9.5$  &  $6.4_{-0.4}^{+0.4}$   &  $-23.91_{-0.58}^{+0.66}$   &  $5.8_{-1.8}^{+2.7}$   &   \\ 
M105  &  E  &  $10.3$  &  $4_{-1}^{+1}$   &  $-24.29_{-0.58}^{+0.66}$   &  $5.2_{-1.6}^{+2.4}$   &   \\ 
M106  &  Sp (bar)  &  $7.2$  &  $0.39_{-0.01}^{+0.01}$   &  $-21.11_{-0.11}^{+0.18}$   &  $2.0_{-0.2}^{+0.3}$   &   \\ 
NGC 0524  &  S0  &  $23.3$  &  $8.3_{-1.3}^{+2.7}$   &  $-23.19_{-0.11}^{+0.18}$   &  $1.1_{-0.1}^{+0.2}$   &   \\ 
NGC 0821  &  E  &  $23.4$  &  $0.39_{-0.09}^{+0.26}$   &  $-24.00_{-0.66}^{+0.88}$   &  $5.3_{-2.3}^{+4.1}$   &   \\ 
NGC 1023  &  S0 (bar)  &  $11.1$  &  $0.42_{-0.04}^{+0.04}$   &  $-22.82_{-0.11}^{+0.18}$   &  $2.1_{-0.3}^{+0.3}$   &   \\ 
NGC 1300  &  Sp (bar)  &  $20.7$  &  $0.73_{-0.35}^{+0.69}$   &  $-22.06_{-0.58}^{+0.66}$   &  $3.8_{-1.2}^{+1.8}$   &   \\ 
NGC 1316  &  merger  &  $18.6$  &  $1.50_{-0.80}^{+0.75}$   &  $-24.89_{-0.58}^{+0.66}$   &  $2.0_{-0.7}^{+1.0}$   &   \\ 
NGC 1332  &  E/S0  &  $22.3$  &  $14_{-2}^{+2}$   &  $-24.89_{-0.66}^{+0.88}$   &  $5.1_{-2.2}^{+3.9}$   &   \\ 
NGC 1374  &  E  &  $19.2$  &  $5.8_{-0.5}^{+0.5}$   &  $-23.68_{-0.11}^{+0.18}$   &  $3.7_{-0.5}^{+0.5}$   &   \\ 
NGC 1399  &  E  &  $19.4$  &  $4.7_{-0.6}^{+0.6}$   &  $-26.43_{-0.11}^{+0.18}$   &  $10.0_{-1.2}^{+1.4}$   &   \\ 
NGC 2273  &  Sp (bar)  &  $28.5$  &  $0.083_{-0.004}^{+0.004}$   &  $-23.00_{-0.58}^{+0.66}$   &  $2.1_{-0.7}^{+1.0}$   &   \\ 
NGC 2549  &  S0 (bar)  &  $12.3$  &  $0.14_{-0.13}^{+0.02}$   &  $-21.25_{-0.11}^{+0.18}$   &  $2.3_{-0.3}^{+0.3}$   &   \\ 
NGC 2778  &  S0 (bar)  &  $22.3$  &  $0.15_{-0.10}^{+0.09}$   &  $-20.80_{-0.58}^{+0.66}$   &  $1.3_{-0.4}^{+0.6}$   &   \\ 
NGC 2787  &  S0 (bar)  &  $7.3$  &  $0.40_{-0.05}^{+0.04}$   &  $-20.11_{-0.58}^{+0.66}$   &  $1.1_{-0.4}^{+0.5}$   &   \\ 
NGC 2974  &  Sp (bar)  &  $20.9$  &  $1.7_{-0.2}^{+0.2}$   &  $-22.95_{-0.58}^{+0.66}$   &  $1.4_{-0.5}^{+0.7}$   &   \\ 
NGC 3079  &  Sp (bar)  &  $20.7$  &  $0.024_{-0.012}^{+0.024}$   &  $-23.01_{-0.58}^{+0.66}$   &  $1.3_{-0.4}^{+0.6}$   &   \\ 
NGC 3091  &  E  &  $51.2$  &  $36_{-2}^{+1}$   &  $-26.28_{-0.11}^{+0.18}$   &  $7.6_{-0.9}^{+1.0}$   &   \\ 
NGC 3115  &  E/S0  &  $9.4$  &  $8.8_{-2.7}^{+10.0}$   &  $-24.22_{-0.11}^{+0.18}$   &  $4.4_{-0.5}^{+0.6}$   &   \\ 
NGC 3227  &  Sp (bar)  &  $20.3$  &  $0.14_{-0.06}^{+0.10}$   &  $-21.76_{-0.58}^{+0.66}$   &  $1.7_{-0.5}^{+0.8}$   &   \\ 
NGC 3245  &  S0 (bar)  &  $20.3$  &  $2.0_{-0.5}^{+0.5}$   &  $-22.43_{-0.11}^{+0.18}$   &  $2.9_{-0.3}^{+0.4}$   &   \\ 
NGC 3377  &  E  &  $10.9$  &  $0.77_{-0.06}^{+0.04}$   &  $-23.49_{-0.58}^{+0.66}$   &  $7.7_{-2.5}^{+3.6}$   &   \\ 
NGC 3384  &  S0 (bar)  &  $11.3$  &  $0.17_{-0.02}^{+0.01}$   &  $-22.43_{-0.11}^{+0.18}$   &  $1.6_{-0.2}^{+0.2}$   &   \\ 
NGC 3393  &  Sp (bar)  &  $55.2$  &  $0.34_{-0.02}^{+0.02}$   &  $-23.48_{-0.58}^{+0.66}$   &  $3.4_{-1.1}^{+1.6}$   &   \\ 
NGC 3414  &  E  &  $24.5$  &  $2.4_{-0.3}^{+0.3}$   &  $-24.35_{-0.11}^{+0.18}$   &  $4.8_{-0.6}^{+0.7}$   &   \\ 
NGC 3489  &  S0/Sp (bar)  &  $11.7$  &  $0.058_{-0.008}^{+0.008}$   &  $-21.13_{-0.58}^{+0.66}$   &  $1.5_{-0.5}^{+0.7}$   &   \\ 
NGC 3585  &  E  &  $19.5$  &  $3.1_{-0.6}^{+1.4}$   &  $-25.52_{-0.58}^{+0.66}$   &  $5.2_{-1.7}^{+2.4}$   &   \\ 
NGC 3607  &  E  &  $22.2$  &  $1.3_{-0.5}^{+0.5}$   &  $-25.36_{-0.58}^{+0.66}$   &  $5.5_{-1.7}^{+2.6}$   &   \\ 
NGC 3608  &  E  &  $22.3$  &  $2.0_{-0.6}^{+1.1}$   &  $-24.50_{-0.58}^{+0.66}$   &  $5.2_{-1.7}^{+2.4}$   &   \\ 
NGC 3842  &  E  &  $98.4$  &  $97_{-26}^{+30}$   &  $-27.00_{-0.11}^{+0.18}$   &  $8.1_{-1.0}^{+1.1}$   &   \\ 
NGC 3998  &  S0 (bar)  &  $13.7$  &  $8.1_{-1.9}^{+2.0}$   &  $-22.32_{-0.66}^{+0.88}$   &  $1.2_{-0.5}^{+0.9}$   &   \\ 
NGC 4026  &  S0 (bar)  &  $13.2$  &  $1.8_{-0.3}^{+0.6}$   &  $-21.58_{-0.66}^{+0.88}$   &  $2.4_{-1.0}^{+1.8}$   &   \\ 
NGC 4151  &  Sp (bar)  &  $20.0$  &  $0.65_{-0.07}^{+0.07}$   &  $-23.40_{-0.58}^{+0.66}$   &  $1.4_{-0.4}^{+0.6}$   &   \\ 
NGC 4261  &  E  &  $30.8$  &  $5_{-1}^{+1}$   &  $-25.72_{-0.58}^{+0.66}$   &  $4.7_{-1.5}^{+2.2}$   &   \\ 
NGC 4291  &  E  &  $25.5$  &  $3.3_{-2.5}^{+0.9}$   &  $-24.05_{-0.58}^{+0.66}$   &  $4.2_{-1.4}^{+2.0}$   &   \\ 
NGC 4388  &  Sp (bar)  &  $17.0$  &  $0.075_{-0.002}^{+0.002}$   &  $-21.26_{-0.66}^{+0.88}$   &  $0.6_{-0.3}^{+0.5}$   &   \\ 
NGC 4459  &  S0  &  $15.7$  &  $0.68_{-0.13}^{+0.13}$   &  $-23.48_{-0.58}^{+0.66}$   &  $3.1_{-1.0}^{+1.5}$   &   \\ 
NGC 4473  &  E  &  $15.3$  &  $1.2_{-0.9}^{+0.4}$   &  $-23.88_{-0.58}^{+0.66}$   &  $2.3_{-0.7}^{+1.1}$   &   \\ 
NGC 4564  &  S0  &  $14.6$  &  $0.60_{-0.09}^{+0.03}$   &  $-22.30_{-0.11}^{+0.18}$   &  $2.6_{-0.3}^{+0.4}$   &   \\ 
NGC 4596  &  S0 (bar)  &  $17.0$  &  $0.79_{-0.33}^{+0.38}$   &  $-22.73_{-0.11}^{+0.18}$   &  $2.7_{-0.3}^{+0.4}$   &   \\ 
\tableline         
\end{tabular}   
\label{tab:sample} 
\end{center}    
\end{table*}    

\begin{table*}                                        
\small                                                
\begin{center}                                        
\begin{tabular}{llllllrll}                           
\tableline                                                
\multicolumn{1}{l}{{\bf Galaxy}} &                   
\multicolumn{1}{l}{{\bf Type}} &                     
\multicolumn{1}{l}{{\bf Distance}} &                 
\multicolumn{1}{l}{{\bf $\bm{M_{\rm BH}}$}} &  
\multicolumn{1}{l}{{\bf $\bm{MAG_{\rm sph}}$}} &  
\multicolumn{1}{l}{{\bf $\bm{n_{\rm sph}^{\rm maj}}$}} \\  
\multicolumn{1}{l}{} &                                
\multicolumn{1}{l}{} &                                
\multicolumn{1}{l}{[Mpc]} &                           
\multicolumn{1}{l}{$[10^8~\rm M_{\odot}]$} &         
\multicolumn{1}{l}{[mag]} &                                
\multicolumn{1}{l}{} \\                             
\multicolumn{1}{l}{(1)} &                             
\multicolumn{1}{l}{(2)} &                             
\multicolumn{1}{l}{(3)} &                             
\multicolumn{1}{l}{(4)} &                             
\multicolumn{1}{l}{(5)} &                             
\multicolumn{1}{l}{(6)} \\  
\tableline                                                
NGC 4697  &  E  &  $11.4$  &  $1.8_{-0.1}^{+0.2}$   &  $-24.82_{-0.66}^{+0.88}$   &  $7.2_{-3.1}^{+5.5}$   &   \\ 
NGC 4889  &  E  &  $103.2$  &  $210_{-160}^{+160}$   &  $-27.54_{-0.11}^{+0.18}$   &  $8.1_{-1.0}^{+1.1}$   &   \\ 
NGC 4945  &  Sp (bar)  &  $3.8$  &  $0.014_{-0.007}^{+0.014}$   &  $-20.96_{-0.58}^{+0.66}$   &  $1.4_{-0.5}^{+0.7}$   &   \\ 
NGC 5077  &  E  &  $41.2$  &  $7.4_{-3.0}^{+4.7}$   &  $-25.45_{-0.11}^{+0.18}$   &  $4.2_{-0.5}^{+0.6}$   &   \\ 
NGC 5128  &  merger  &  $3.8$  &  $0.45_{-0.10}^{+0.17}$   &  $-23.89_{-0.66}^{+0.88}$   &  $1.2_{-0.5}^{+0.9}$   &   \\ 
NGC 5576  &  E  &  $24.8$  &  $1.6_{-0.4}^{+0.3}$   &  $-24.44_{-0.11}^{+0.18}$   &  $3.3_{-0.4}^{+0.5}$   &   \\ 
NGC 5845  &  S0  &  $25.2$  &  $2.6_{-1.5}^{+0.4}$   &  $-22.96_{-0.66}^{+0.88}$   &  $2.5_{-1.1}^{+1.9}$   &   \\ 
NGC 5846  &  E  &  $24.2$  &  $11_{-1}^{+1}$   &  $-25.81_{-0.58}^{+0.66}$   &  $6.4_{-2.1}^{+3.0}$   &   \\ 
NGC 6251  &  E  &  $104.6$  &  $5_{-2}^{+2}$   &  $-26.75_{-0.11}^{+0.18}$   &  $6.8_{-0.8}^{+0.9}$   &   \\ 
NGC 7052  &  E  &  $66.4$  &  $3.7_{-1.5}^{+2.6}$   &  $-26.32_{-0.11}^{+0.18}$   &  $4.2_{-0.5}^{+0.6}$   &   \\ 
NGC 7619  &  E  &  $51.5$  &  $25_{-3}^{+8}$   &  $-26.35_{-0.58}^{+0.66}$   &  $5.3_{-1.7}^{+2.5}$   &   \\ 
NGC 7768  &  E  &  $112.8$  &  $13_{-4}^{+5}$   &  $-26.90_{-0.58}^{+0.66}$   &  $8.4_{-2.7}^{+3.9}$   &   \\ 
UGC 03789  &  Sp (bar)  &  $48.4$  &  $0.108_{-0.005}^{+0.005}$   &  $-22.77_{-0.66}^{+0.88}$   &  $1.9_{-0.8}^{+1.4}$   &   \\ 
\tableline         
\end{tabular}   
\tablecomments{\emph{Column (1)}: Galaxy name. 
\emph{Column (2)}: Morphological type (E=elliptical, S0=lenticular, Sp=spiral, merger). 		       The morphological classification of four galaxies is uncertain (E/S0 or S0/Sp). 		       The presence of a bar is indicated. 
\emph{Column (3)}: Distance. 
\emph{Column (4)}: Black hole mass. 
\emph{Column (5)}: Absolute $3.6\rm~\mu m$ spheroid magnitude. \ 
\emph{Column (6)}: Spheroid major-axis S\'ersic index. \ 
Spheroid magnitudes and S\'ersic indices come from our state-of-the-art multicomponent galaxy decompositions (\emph{Paper I}), 		       
which include bulge, disks, bars, spiral arms, rings, halo, extended or unresolved nuclear source and partially depleted core,                        
and that -- for the first time -- were checked to be consistent with the galaxy kinematics. 		       
The uncertainties were estimated with a method that takes into account systematic errors, which are typically not considered by popular 2D fitting codes. } 
\end{center}    
\end{table*}

\section{Analysis and results}
\label{sec:anal}
\begin{figure}[h]
\begin{center}
\includegraphics[width=\columnwidth, trim = 20 0 220 0]{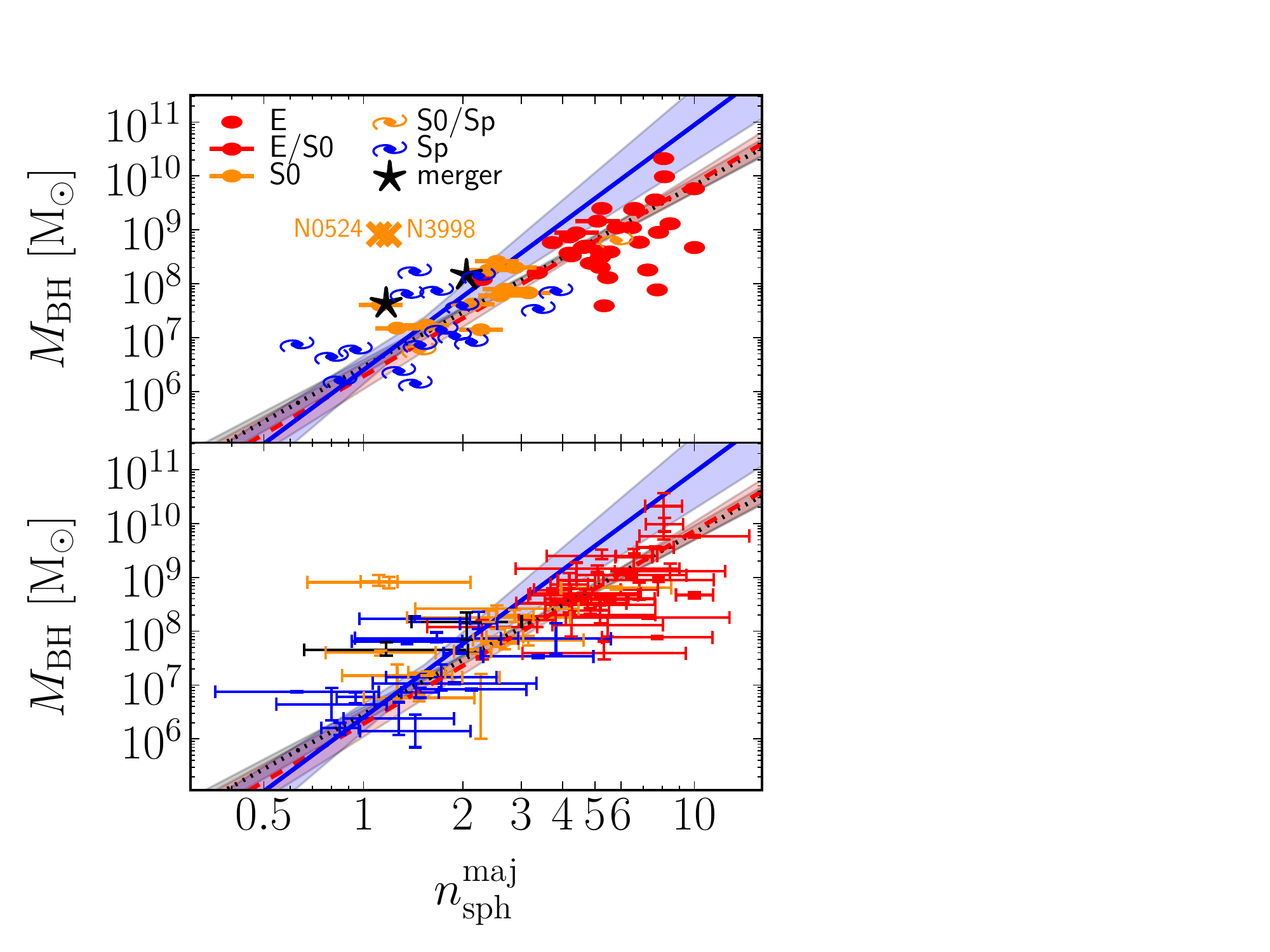}
\caption{Black hole mass plotted against spheroid S\'ersic index measured along the galaxy major-axis. 
The top and bottom panels show the data points and their error bars, respectively.  
Symbols are coded according to the galaxy morphological type (see legend). 
The orange crosses denote two lenticular galaxies (NGC 0524 and NGC 3998) which were excluded from the linear regression analysis 
(see Section \ref{sec:anal}). 
The black dotted line is the FITEXY bisector linear regression for all ($64$) galaxies, 
with the gray shaded area denoting its $1\sigma$ uncertainty. 
The red dashed line is the FITEXY bisector linear regression for the $45-2=43$  early-type galaxies (E+S0), 
with the red shaded area denoting its $1\sigma$ uncertainty. 
The blue solid line is the FITEXY bisector linear regression for the bulges of the 17 late-type (Sp) galaxies, 
with the blue shaded area denoting its $1\sigma$ uncertainty. 
The linear regressions for early- and late-type galaxies are consistent with each other within their $1\sigma$ uncertainties.  
The error bars in the bottom panel have the same color coding as the symbols in the top panel. 
}
\label{fig:mbhn}
\end{center}
\end{figure}

As in \emph{Paper II}, a linear regression analysis of the $L_{\rm sph} - n_{\rm sph}$ 
(Table \ref{tab:lregLn} and Figure \ref{fig:magn})
and $M_{\rm BH} - n_{\rm sph}$ (Table \ref{tab:lregMn} and Figure \ref{fig:mbhn}) diagrams 
was performed using three different routines: 
the BCES code from \cite{akritasbershady1996}, 
the FITEXY routine \citep{press1992}, as modified by \cite{tremaine2002}, 
and the Bayesian estimator {\tt linmix\_err} \citep{linmixerr}.
All of these three routines take into account the intrinsic scatter, 
but only the FITEXY and the {\tt linmix\_err} codes allow one to quantify it.
\cite{tremaine2002} cautioned that 
the BCES estimator becomes ineffective when the dataset contains at least one low-precision measurement -- 
regardless of how many high-precision measurements are in the sample -- 
and tends to be biased in case of low number statistics, 
or if the mean square of the uncertainties associated to the independent variable is comparable to 
the variance of the distribution of the independent variable. 
According to the results from the Monte Carlo Markov Chain simulations of \cite{tremaine2002} and \cite{novak2006}, 
these problems can be overcome with the use of the modified FITEXY routine. 
\cite{park2012} also concluded that the modified FITEXY routine performs better and returns less biased results than the BCES estimator, 
and noted that the the modified FITEXY routine is computationally less intensive than the Bayesian technique {\tt linmix\_err}. 
Given that at least one of our subsamples (the lenticular galaxies) has a small size 
and that the uncertainties associated to $n_{\rm sph}$ are relatively large 
compared to the range spanned by the $n_{\rm sph}$ values for most of our subsamples, 
we put more trust in the results obtained with the modified FITEXY routine 
and throughout the text we quote only those. \\ 
We report both symmetrical and nonsymmetrical linear regressions.  
Symmetrical regressions are meant to be compared with theoretical expectations, 
whereas nonsymmetrical forward ($Y|X$) regressions -- which minimize the scatter in the $Y$ direction -- 
allow one to predict the value of the observable $Y$ with the best possible precision. \\
We searched for extreme outliers in both the $L_{\rm sph} - n_{\rm sph}$ and $M_{\rm BH} - n_{\rm sph}$ diagrams, 
and found that in our $L_{\rm sph} - n_{\rm sph}$ plot there are no $3\sigma$ outliers, 
whereas in our $M_{\rm BH} - n_{\rm sph}$ plot 
the lenticular galaxies NGC 0524 and NGC 3998 reside more than $3\sigma$ from the bisector linear regression for all galaxies. 
These two galaxies have therefore been excluded from the rest of the analysis. 

\subsection{$L_{\rm sph} - n_{\rm sph}$}
\begin{figure}[h]
\begin{center}
\includegraphics[width=\columnwidth, trim = 20 0 220 0]{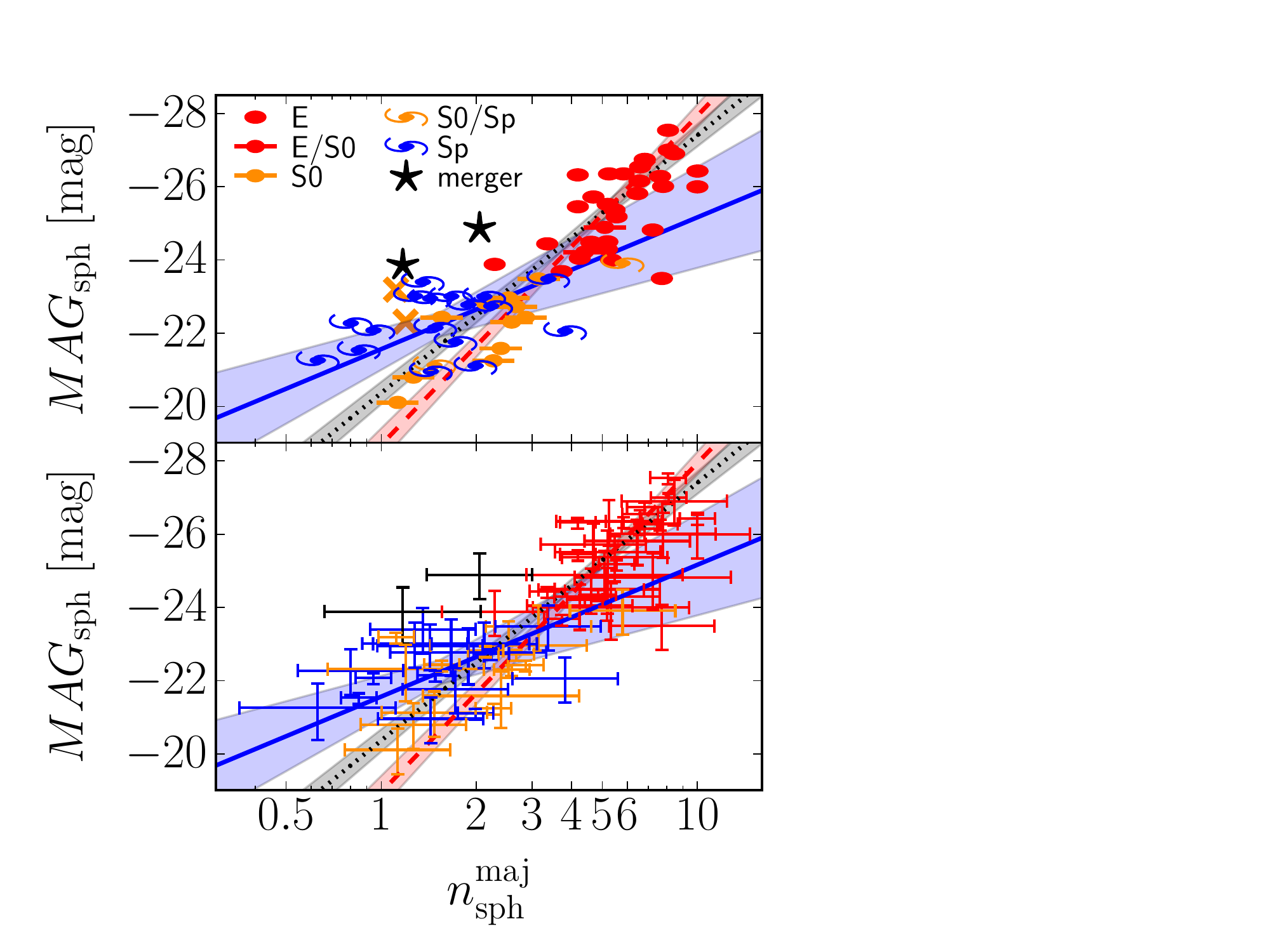}
\caption{Spheroid absolute magnitude (at $3.6\rm~\mu m$) plotted against spheroid S\'ersic index 
measured along the galaxy major-axis. 
The top and bottom panels show the data points and their error bars, respectively.  
Symbols are coded according to the galaxy morphological type (see legend in the top panel).
The orange crosses denote two lenticular galaxies (NGC 0524 and NGC 3998) which were excluded from the linear regression analysis 
(see Section \ref{sec:anal}). 
The black dotted line is the FITEXY bisector linear regression for all ($64$) galaxies, 
with the gray shaded area denoting its $1\sigma$ uncertainty. 
The red dashed line is the FITEXY bisector linear regression for the $45-2=43$  early-type galaxies (E+S0), 
with the red shaded area denoting its $1\sigma$ uncertainty. 
The shallower blue solid line is the FITEXY bisector linear regression for the bulges of the 17 late-type (Sp) galaxies, 
with the blue shaded area denoting its $1\sigma$ uncertainty. 
The error bars in the bottom panel have the same color coding as the symbols in the top panel. 
}
\label{fig:magn}
\end{center}
\end{figure}

Following \cite{graham2001bulges}, 
who showed that the $L_{\rm sph} - n_{\rm sph}$ relation is different for elliptical galaxies and the bulges of disk galaxies (S0+Sp), 
\cite{savorgnan2013} re-analyzed the data from \cite{grahamguzman2003} and \cite{grahamworley2008} 
and obtained two separate $L_{\rm sph} - n_{\rm sph}$ linear regressions for elliptical galaxies and the bulges of disk galaxies 
(in the B- and K-band, respectively). 
At the time, the $L_{\rm sph} - n_{\rm sph}$ datasets from \cite{grahamguzman2003} and \cite{grahamworley2008} were of the best quality available 
to investigate the $L_{\rm sph} - n_{\rm sph}$ relation for different galaxy morphological types. 
However, these datasets were not obtained from a homogeneous analysis, 
but they were a collection of results taken from various past bulge/disk decomposition studies. 
Here we re-investigate the $L_{\rm sph} - n_{\rm sph}$ diagram (Figure \ref{fig:magn}) using only our high-quality dataset. 
Our spheroid luminosities and S\'ersic indices were obtained from accurate multicomponent decompositions,  
performed in a consistent manner using the $3.6\rm~\mu m$ band, which is less affected by dust extinction than the K-band. \\
\cite{grahamworley2008} presented a single $L_{\rm sph} - n_{\rm sph}$ correlation for the bulges of disk galaxies (S0+Sp).
However, using our dataset, 
we fit the $L_{\rm sph} - n_{\rm sph}$ relation for elliptical, lenticular and spiral galaxies separately, 
and found that the values of the slope and intercept for the lenticular galaxies 
are not consistent within the errors with those for the spiral galaxies, 
but are consistent within the errors with those for the elliptical galaxies.  
Given this, we conclude that in the $L_{\rm sph} - n_{\rm sph}$ diagram 
elliptical and lenticular galaxies form together a single (\emph{early-type}) sequence, 
whereas the combination of lenticular and spiral galaxies do not. 
According to the modified FITEXY routine, 
early-type galaxies\footnote{Using the BCES estimator, 
\cite{savorgnan2013} re-analyzed the dataset from \cite{grahamguzman2003} 
and obtained $L_{\rm sph} \propto n_{\rm sph}^{3.60 \pm 0.19}$ for the elliptical galaxies only. 
This result is in excellent agreement with the BCES linear regression obtained here for the early-type galaxies 
($L_{\rm sph} \propto n_{\rm sph}^{3.89 \pm 0.42}$) 
and, remarkably, it is exactly the same proportionality obtained here for the early-type galaxies with the modified FITEXY routine. } 
follow $L_{\rm sph} \propto n_{\rm sph}^{3.60 \pm 0.19}$, 
whereas late-type galaxies follow a shallower $L_{\rm sph} \propto n_{\rm sph}^{1.44 \pm 0.52}$ sequence. \\
Because the log-slopes of the correlations for early- and late-type galaxies are not consistent with each other 
within their $1\sigma$ uncertainties, 
our quantitative linear regression analysis suggests that 
the $L_{\rm sph} - n_{\rm sph}$ diagram is better described with a four-parameter model 
(two separate power-laws) 
rather than with a two-parameter model (single power-law). 
In addition, the relative quality of these two statistical models can be independently assessed 
using the Akaike Information Criterion ($AIC$, \citealt{akaike1974}). 
$AIC$ is a trade-off between the statistical significance of a fit and the complexity of the model used. 
It benefits from the goodness of a fit, 
but at the same time is also penalized by the number of parameters of the model, 
hence it discourages overfitting. 
The $AICc$ is a variation of the $AIC$ that takes into account a correction for finite sample sizes: 
\begin{equation}
AICc = 2k - 2 \ln(\mathcal{L}) + \frac{2k (k+1)}{n-k-1} ,
\end{equation}
where $k$ is the number of parameter of the model, 
$n$ is the sample size, 
and $\mathcal{L}$ is the maximum value of the likelihood function for the model. 
Within a set of candidate models for a given dataset, 
the best model has the smallest $AICc$ value. 
Using our $L_{\rm sph} - n_{\rm sph}$ dataset, 
the $AICc$ value for a double power-law model is a factor of $3/4$ smaller 
than the $AICc$ value for a single power-law model.

\subsection{$M_{\rm BH} - n_{\rm sph}$}
\cite{grahamscott2013} presented two different $M_{\rm BH} - L_{\rm sph}$ relations for S\'ersic and core-S\'ersic 
spheroids\footnote{Core-S\'ersic spheroids have partially depleted cores relative to their outer S\'ersic light profile, 
whereas S\'ersic spheroids have no central deficit of stars. } \citep{graham2003coresersicmodel,trujillo2004coresersicmodel}.
However, in \emph{Paper II} we found that the slopes of the $M_{\rm BH} - L_{\rm sph}$ correlations for S\'ersic and core-S\'ersic spheroids 
are consistent with each other within their $1\sigma$ uncertainties, 
which prevented us from considering them as two separate sequences. 
On the other hand, our analysis showed that early- and late-type galaxies follow two different $M_{\rm BH} - L_{\rm sph}$ relations. 
Given that early- and late-type galaxies define two separate sequences in both the $L_{\rm sph} - n_{\rm sph}$ 
and $M_{\rm BH} - L_{\rm sph}$ diagrams, 
we investigate substructure in the $M_{\rm BH} - n_{\rm sph}$ diagram. 
Using the results from the modified FITEXY routine, 
we know that the early-type galaxies follow $M_{\rm BH} \propto L_{\rm sph}^{1.03 \pm 0.11}$ 
and $L_{\rm sph} \propto n_{\rm sph}^{3.60 \pm 0.19}$, 
therefore we expect to find $M_{\rm BH} \propto n_{\rm sph}^{3.69 \pm 0.44}$; 
this prediction is in excellent agreement with the observed log-slope of $3.58 \pm 0.27$ obtained here. 
On the other hand, late-type galaxies follow $M_{\rm BH} \propto L_{\rm sph}^{2.58 \pm 1.06}$ 
and $L_{\rm sph} \propto n_{\rm sph}^{1.44 \pm 0.52}$, 
from which one can predict $M_{\rm BH} \propto n_{\rm sph}^{3.72 \pm 2.03}$; 
this is consistent with the observed log-slope of $4.55 \pm 0.66$. 
The Bayesian estimator {\tt linmix\_err} returns consistent results: 
a log-slope of $3.44 \pm 0.33$ for the early-type galaxies 
and a log-slope of $4.12 \pm 1.07$ for the late-type galaxies. 
Regardless of the linear regression routine used, 
the values of the slope and intercept for the early- and late-type galaxies 
are consistent with each other within their $1\sigma$ uncertainties\footnote{In effect, 
considering the results of the modified FITEXY routine, 
the slopes of the relations for early- and late-type galaxies 
are only marginally consistent with each other within their $1\sigma$ uncertainties. 
However, the same slopes obtained with Bayesian estimator {\tt linmix\_err} 
are fully consistent with each other within their $1\sigma$ uncertainties. }.
Our analysis shows that the early- and late-type galaxies do not follow two separate trends 
in the $M_{\rm BH} - n_{\rm sph}$ diagram, 
i.e.~we do not identify any significant substructure based on the galaxy morphological type. 
The $AICc$ value for a single power-law model is a factor of $2/3$ smaller 
than the $AICc$ value for a double power-law model.  \\
The symmetrical linear regression for all galaxies obtained with the modified FITEXY routine is: 
\begin{equation*}
\log \biggl( \frac{M_{\rm BH}}{\rm M_\odot} \biggr) = (8.15 \pm 0.06) + (3.37 \pm 0.15) \bigl(\log n_{\rm sph} - 0.50 \bigr) . 
\end{equation*}
\\

We have seen that the early- and late-type galaxies in the $M_{\rm BH} - n_{\rm sph}$ diagram 
can be fit (together) with a single power-law, 
whereas they follow two different correlations in the $M_{\rm BH} - L_{\rm sph}$ diagram (\emph{Paper II}). 
We now want to compare the amount of intrinsic scatter in these two plots. 
In Table \ref{tab:intsc} we report the values of the intrinsic scatter 
in the $M_{\rm BH} - n_{\rm sph}$ and $M_{\rm BH} - L_{\rm sph}$ diagrams 
for all, early- and late-type galaxies, 
obtained with the modified FITEXY routine and the Bayesian estimator {\tt linmix\_err}. 
When considering all galaxies, irrespective of their morphological type, 
the intrinsic scatter of the $M_{\rm BH} - n_{\rm sph}$ relation is smaller than that of the $M_{\rm BH} - L_{\rm sph}$ relation. 
However, this is obviously not a fair comparison, 
because of the different nature of the $M_{\rm BH} - n_{\rm sph}$ and $M_{\rm BH} - L_{\rm sph}$ correlations 
(single and double power-law, respectively). 
One can obtain more informative results by considering early- and late-type galaxies separately. 
For the early-type galaxies, the intrinsic scatter of the $M_{\rm BH} - n_{\rm sph}$ relation is consistent\footnote{Looking 
at the results obtained with the modified FITEXY routine, 
the values of the intrinsic scatter are only marginally consistent with each other, 
but looking at the results obtained with the Bayesian estimator {\tt linmix\_err}, 
the values of the intrinsic scatter are fully consistent with each other. } 
with that of the $M_{\rm BH} - L_{\rm sph}$ relation (within their $1\sigma$ uncertainties). 
For the late-type galaxies, the intrinsic scatter of the $M_{\rm BH} - n_{\rm sph}$ relation is consistent with 
that of the $M_{\rm BH} - L_{\rm sph}$ relation, 
except for the inverse ($X|Y$) regression obtained with the modified FITEXY routine. \\
In passing, we note that the values of the intrinsic scatter of the $M_{\rm BH} - n_{\rm sph}$ relation 
are systematically smaller -- although consistent within the errors -- 
than the corresponding values of the intrinsic scatter of the $M_{\rm BH} - L_{\rm sph}$ relation.  
In addition, the values of the intrinsic scatter returned by the modified FITEXY routine are systematically smaller than 
those output by the Bayesian estimator {\tt linmix\_err}.

\section{Conclusions} 
The S\'ersic index of a galaxy's spheroidal component, $n_{\rm sph}$, constitutes a direct measure 
of its central radial concentration of stars. 
After \cite{graham2001bulges} proved that the central black hole mass, $M_{\rm BH}$, 
correlates with the stellar light concentration of a galaxy's spheroidal component, 
\cite{grahamdriver2007} presented and analyzed for the first time a tight $M_{\rm BH} - n_{\rm sph}$ correlation 
using a sample of 27 elliptical and disk galaxies for which they had performed photometric bulge/disk decomposition. 
The $M_{\rm BH} - n_{\rm sph}$ correlation can be predicted from the combination of two well known scaling relations 
involving the spheroid luminosity, $L_{\rm sph}$: 
the $M_{\rm BH} - L_{\rm sph}$ (e.g.~\citealt{kormendyrichstone1995,magorrian1998,marconihunt2003,haringrix2004}) 
and the $L_{\rm sph} - n_{\rm sph}$ (e.g.~\citealt{youngcurrie1994,jerjen2000,grahamguzman2003}). 
However, upon independently attempting photometric multicomponent decompositions 
for galaxy samples that were similar to that used by \cite{grahamdriver2007}, 
three subsequent studies \citep{sani2011,vika2012,beifiori2012} failed to recover a statistically significant $M_{\rm BH} - n_{\rm sph}$ relation. 
\cite{savorgnan2013} collected and compared the S\'ersic index measurements obtained by \cite{grahamdriver2007}, 
\cite{sani2011}, \cite{vika2012} and \cite{beifiori2012}, 
and argued that the discrepancies were due to inaccurate galaxy decompositions. \\
Moved by a urgent need to re-investigate and refine several black hole mass scaling relations, 
we performed state-of-the-art photometric multicomponent decompositions 
(i.e.~bulge, disks, bars, spiral arms, rings, halo, nucleus, depleted core, etc.) 
for 66 galaxies with a dynamical measurement of their black hole mass (\emph{Paper I}). 
We carefully measured the S\'ersic index of each galaxy's spheroidal component 
and estimated its associated uncertainty with a method that takes into account statistical and systematic errors. 
Our analysis shows that early- (elliptical + lenticular) and late-type (spiral) galaxies define two different correlations 
in the $L_{\rm sph} - n_{\rm sph}$ diagram, 
whereas they reunite in a single sequence in the $M_{\rm BH} - n_{\rm sph}$ diagram. 
With the current dataset, we measured an amount of intrinsic scatter in the $M_{\rm BH} - n_{\rm sph}$ diagram systematically smaller, 
but still consistent within the errors, with that observed in the the $M_{\rm BH} - L_{\rm sph}$ diagram (\emph{Paper II}). \\ 
Our results suggest that the black hole mass is intimately connected to the spheroid central concentration of stars, 
which reflects the inner gradient of the spheroid gravitational potential. 
Besides conferring the spheroid S\'ersic index a predictive power to infer the black hole mass 
from a galaxy's image only (even photometrically uncalibrated), 
the $M_{\rm BH} - n_{\rm sph}$ correlation should become a fundamental ingredient in semi-analytic models and simulations of galaxy formation.

\begin{table*}
\centering
\caption{Linear regression analysis of the $L_{\rm sph} - n_{\rm sph}$ diagram.}
\begin{tabular}{llccccc}
\tableline
\tableline
{\bf Subsample (size)} & {\bf Regression} & $\boldsymbol \alpha$ & $\boldsymbol \beta$ & $\boldsymbol{\langle \log n_{\rm sph}^{\rm maj} \rangle}$ & $\boldsymbol \epsilon$ & $\boldsymbol \Delta$ \\ 
\tableline 
\\
 & \multicolumn{6}{l}{$MAG_{\rm sph}/{\rm [mag]} = \alpha + \beta \bigl(\log n_{\rm sph}^{\rm maj} - \langle \log n_{\rm sph}^{\rm maj} \rangle \bigr)$} \\ [0.5em]
All (64)               & BCES $(Y|X)$               & $-23.89 \pm 0.15$ & $-6.90 \pm 0.74$ & $0.50$ & $-$ & $1.22$ \\
                       & mFITEXY $(Y|X)$            & $-23.91 \pm 0.13$ & $-6.63 \pm 0.45$ & $0.50$ & $0.59^{+0.16}_{-0.11}$ & $1.01$ \\
                       & {\tt linmix\_err} $(Y|X)$  & $-23.89 \pm 0.14$ & $-6.34 \pm 0.57$ & $0.50$ & $0.74 \pm 0.13$ & $1.14$ \\ [0.5em]
                       & BCES $(X|Y)$               & $-23.89 \pm 0.15$ & $-6.75 \pm 0.52$ & $0.50$ & $-$ & $1.20$ \\
                       & mFITEXY $(X|Y)$            & $-23.89 \pm 0.14$ & $-7.49 \pm 0.53$ & $0.50$ & $0.62^{+0.18}_{-0.12}$ & $1.32$ \\
                       & {\tt linmix\_err} $(X|Y)$  & $-23.90 \pm 0.16$ & $-7.49 \pm 0.62$ & $0.50$ & $0.80 \pm 0.16$ & $1.32$ \\ [0.5em]
                       & BCES Bisector              & $-23.89 \pm 0.15$ & $-6.83 \pm 0.58$ & $0.50$ & $-$ & $1.21$ \\
                       & mFITEXY Bisector           & $-23.90 \pm 0.13$ & $-7.04 \pm 0.35$ & $0.50$ & $-$ & $1.24$ \\
                       & {\tt linmix\_err} Bisector & $-23.89 \pm 0.15$ & $-6.87 \pm 0.42$ & $0.50$ & $-$ & $1.21$ \\ [0.5em]

Elliptical (E) (30)    & BCES $(Y|X)$		    & $-25.46 \pm 1.12$ & $38.47 \pm 114.45$ & $0.76$ & $-$ & $6.37$ \\
		       & mFITEXY $(Y|X)$	    & $-25.74 \pm 0.18$ & $-9.74 \pm 1.59$ & $0.76$ & $0.24^{+0.32}_{-0.24}$ & $0.94$ \\
		       & {\tt linmix\_err} $(Y|X)$  & $-25.65 \pm 0.21$ & $-7.87 \pm 2.15$ & $0.76$ & $0.61 \pm 0.22$ & $1.06$ \\ [0.5em]
		       & BCES $(X|Y)$		    & $-25.46 \pm 0.23$ & $-10.73 \pm 3.21$ & $0.76$ & $-$ & $1.29$ \\
		       & mFITEXY $(X|Y)$	    & $-25.74 \pm 0.20$ & $-10.42 \pm 1.79$ & $0.76$ & $0.22^{+0.38}_{-0.22}$ & $1.29$ \\
		       & {\tt linmix\_err} $(X|Y)$  & $-25.72 \pm 0.28$ & $-10.92 \pm 2.70$ & $0.76$ & $0.73 \pm 0.34$ & $1.33$ \\ [0.5em]
		       & BCES Bisector  	    & $-25.46 \pm 0.20$ & $0.03 \pm 0.05$ & $0.76$ & $-$ & $1.14$ \\
		       & mFITEXY Bisector	    & $-25.74 \pm 0.19$ & $-10.07 \pm 1.19$ & $0.76$ & $-$ & $1.26$ \\
		       & {\tt linmix\_err} Bisector & $-25.68 \pm 0.25$ & $-9.15 \pm 1.74$ & $0.76$ & $-$ & $1.16$ \\ [0.5em]

Lenticular (S0) (11)   & BCES $(Y|X)$		    & $-22.08 \pm 1.66$ & $33.52 \pm 98.87$ & $0.33$ & $-$ & $6.09$ \\
		       & mFITEXY $(Y|X)$	    & $-22.11 \pm 0.24$ & $-6.31 \pm 2.45$ & $0.33$ & $0.42^{+0.28}_{-0.17}$ & $0.71$ \\
		       & {\tt linmix\_err} $(Y|X)$  & $$ & $$ & $0.33$ & $$ & $$ \\ [0.5em]
		       & BCES $(X|Y)$		    & $-22.08 \pm 0.19$ & $-6.83 \pm 1.16$ & $0.33$ & $-$ & $0.71$ \\
		       & mFITEXY $(X|Y)$	    & $-21.94 \pm 0.44$ & $-13.16 \pm 7.91$ & $0.33$ & $0.61^{+0.60}_{-0.56}$ & $1.39$ \\
		       & {\tt linmix\_err} $(X|Y)$  & $$ & $$ & $0.33$ & $$ & $$ \\ [0.5em]
		       & BCES Bisector  	    & $-22.08 \pm 0.30$ & $0.06 \pm 0.05$ & $0.33$ & $-$ & $1.09$ \\
		       & mFITEXY Bisector	    & $-22.05 \pm 0.35$ & $-8.55 \pm 2.79$ & $0.33$ & $-$ & $0.84$ \\
		       & {\tt linmix\_err} Bisector & $$ & $$ & $0.33$ & $-$ & $$ \\ [0.5em]

Spiral (Sp) (17)       & BCES $(Y|X)$		    & $-22.33 \pm 0.26$ & $-5.31 \pm 5.83$ & $0.18$ & $-$ & $1.15$ \\
		       & mFITEXY $(Y|X)$	    & $-22.22 \pm 0.19$ & $-2.17 \pm 0.98$ & $0.18$ & $0.53^{+0.24}_{-0.13}$ & $0.72$ \\
		       & {\tt linmix\_err} $(Y|X)$  & $-22.26 \pm 0.24$ & $-1.53 \pm 1.88$ & $0.18$ & $0.71 \pm      0.22$ & $0.78$ \\ [0.5em]
		       & BCES $(X|Y)$		    & $-22.33 \pm 0.26$ & $-5.19 \pm 3.77$ & $0.18$ & $-$ & $1.13$ \\
		       & mFITEXY $(X|Y)$	    & $-22.28 \pm 0.44$ & $-9.08 \pm 5.31$ & $0.51$ & $1.12^{+0.54}_{-0.31}$ & $1.83$ \\
		       & {\tt linmix\_err} $(X|Y)$  & $-22.24 \pm 0.71$ & $-11.12 \pm 13.59$ & $0.18$ & $1.95 \pm 2.47$ & $2.24$ \\ [0.5em]
		       & BCES Bisector  	    & $-22.33 \pm 0.26$ & $-5.25 \pm 3.38$ & $0.18$ & $-$ & $1.14$ \\
		       & {\bf mFITEXY Bisector}	    & $\boldsymbol{-22.23 \pm 0.33}$ & $\boldsymbol{-3.60 \pm 1.29}$ & $\boldsymbol{0.18}$ & $-$ & $\boldsymbol{0.92}$ \\
		       & {\tt linmix\_err} Bisector & $-22.25 \pm 0.53$ & $-2.88 \pm 2.66$ & $0.18$ & $-$ & $0.84$ \\ [0.5em]

\tableline 
\tableline
\end{tabular}
\end{table*}

\begin{table*}
\centering
\caption{Linear regression analysis of the $L_{\rm sph} - n_{\rm sph}$ diagram.}
\begin{tabular}{llccccc}
\tableline
\tableline
{\bf Subsample (size)} & {\bf Regression} & $\boldsymbol \alpha$ & $\boldsymbol \beta$ & $\boldsymbol{\langle \log n_{\rm sph}^{\rm maj} \rangle}$ & $\boldsymbol \epsilon$ & $\boldsymbol \Delta$ \\ 
\tableline 
\\
Early-type (E+S0) (43) & BCES $(Y|X)$		    & $-24.55 \pm 0.22$ & $-11.84 \pm 2.29$ & $0.64$ & $-$ & $1.50$ \\
		       & mFITEXY $(Y|X)$	    & $-24.74 \pm 0.14$ & $-8.86 \pm 0.66$ & $0.51$ & $0.27^{+0.20}_{-0.27}$ & $0.87$ \\
		       & {\tt linmix\_err} $(Y|X)$  & $-24.70 \pm 0.17$ & $-8.28 \pm 0.87$ & $0.64$ & $0.58 \pm 0.17$ & $0.98$ \\ [0.5em]
		       & BCES $(X|Y)$		    & $-24.55 \pm 0.14$ & $-8.25 \pm 0.63$ & $0.64$ & $-$ & $0.96$ \\
		       & mFITEXY $(X|Y)$	    & $-24.74 \pm 0.14$ & $-9.13 \pm 0.68$ & $0.64$ & $0.23^{+0.25}_{-0.23}$ & $1.08$ \\
		       & {\tt linmix\_err} $(X|Y)$  & $-24.73 \pm 0.18$ & $-9.08 \pm 0.87$ & $0.64$ & $0.60 \pm 0.21$ & $1.07$ \\ [0.5em]
		       & BCES Bisector  	    & $-24.55 \pm 0.17$ & $-9.73 \pm 1.05$ & $0.64$ & $-$ & $1.14$ \\
		       & mFITEXY Bisector	    & $\boldsymbol{-24.74 \pm 0.14}$ & $\boldsymbol{-8.99 \pm 0.48}$ & $\boldsymbol{0.64}$ & $-$ & $\boldsymbol{1.06}$ \\
		       & {\tt linmix\_err} Bisector & $-24.72 \pm 0.17$ & $-8.66 \pm 0.63$ & $0.64$ & $-$ & $1.02$ \\ [0.5em]

Bulge (S0+Sp) (30)     & BCES $(Y|X)$		    & $-22.25 \pm 0.20$ & $-5.88 \pm 3.06$ & $0.26$ & $-$ & $1.16$ \\
		       & mFITEXY $(Y|X)$	    & $-22.19 \pm 0.14$ & $-2.99 \pm 0.73$ & $0.26$ & $0.52^{+0.18}_{-0.10}$ & $0.75$ \\
		       & {\tt linmix\_err} $(Y|X)$  & $-22.20 \pm 0.17$ & $-2.48 \pm 1.21$ & $0.26$ & $0.67 \pm 0.15$ & $0.83$ \\ [0.5em]
		       & BCES $(X|Y)$		    & $-22.25 \pm 0.20$ & $-5.85 \pm 1.83$ & $0.26$ & $-$ & $1.15$ \\
		       & mFITEXY $(X|Y)$	    & $-22.17 \pm 0.25$ & $-7.65 \pm 2.43$ & $0.26$ & $0.87^{+0.30}_{-0.18}$ & $1.46$ \\
		       & {\tt linmix\_err} $(X|Y)$  & $-22.16 \pm 0.31$ & $-7.80 \pm 3.89$ & $0.26$ & $1.18 \pm 0.65$ & $1.48$ \\ [0.5em]
		       & BCES Bisector  	    & $-22.25 \pm 0.20$ & $-5.87 \pm 2.06$ & $0.26$ & $-$ & $1.16$ \\
		       & mFITEXY Bisector	    & $-22.18 \pm 0.20$ & $-4.34 \pm 0.84$ & $0.26$ & $-$ & $0.96$ \\
		       & {\tt linmix\_err} Bisector & $-22.19 \pm 0.25$ & $-3.83 \pm 1.39$ & $0.26$ & $-$ & $0.91$ \\ [1.0em]

\tableline 
\tableline
\end{tabular}
\label{tab:lregLn} 
\tablecomments{For each subsample, we indicate $\langle \log n_{\rm sph} \rangle$, its average value of spheroid S\'ersic index. 
In the last two columns, we report $\epsilon$, the intrinsic scatter, and $\Delta$, the total rms scatter in the $L_{\rm sph}$ direction. 
The lenticular galaxies NGC 0524 and NGC 3998 were excluded from the linear regression analysis (see Section \ref{sec:anal}). 
Both the early- and late-type subsamples do not contain the two galaxies classified as S0/Sp and the two galaxies classified as mergers (45+17=66-2-2). }
\end{table*}

\begin{table*}
\centering
\caption{Linear regression analysis of the $M_{\rm BH} - n_{\rm sph}$ diagram.}
\begin{tabular}{llccccc}
\tableline
\tableline
{\bf Subsample (size)} & {\bf Regression} & $\boldsymbol \alpha$ & $\boldsymbol \beta$ & $\boldsymbol{\langle \log n_{\rm sph}^{\rm maj} \rangle}$ & $\boldsymbol \epsilon$ & $\boldsymbol \Delta$ \\ 
\tableline 
\\
 & \multicolumn{6}{l}{$\log \bigl( M_{\rm BH}/{\rm [M_\odot]} \bigr) = \alpha + \beta \bigl(\log n_{\rm sph}^{\rm maj} - \langle \log n_{\rm sph}^{\rm maj} \rangle \bigr)$} \\ [0.5em]
 All (64)		& BCES $(Y|X)$  	     & $8.14 \pm 0.07$ & $3.49 \pm 0.36$ & $0.50$ & $-$ & $0.61$ \\
 			& mFITEXY $(Y|X)$	     & $8.15 \pm 0.06$ & $3.26 \pm 0.21$ & $0.50$ & $0.22^{+0.10}_{-0.07}$ & $0.46$ \\
 			& {\tt linmix\_err} $(Y|X)$  & $8.15 \pm 0.06$ & $3.17 \pm 0.24$ & $0.50$ & $0.28 \pm 0.07$ & $0.56$ \\ [0.5em]
 			& BCES $(X|Y)$  	     & $8.14 \pm 0.08$ & $3.52 \pm 0.25$ & $0.50$ & $-$ & $0.61$ \\
 			& mFITEXY $(X|Y)$	     & $8.15 \pm 0.06$ & $3.49 \pm 0.23$ & $0.50$ & $0.23^{+0.10}_{-0.07}$ & $0.61$ \\
 			& {\tt linmix\_err} $(X|Y)$  & $8.15 \pm 0.07$ & $3.49 \pm 0.26$ & $0.50$ & $0.29 \pm 0.08$ & $0.61$ \\ [0.5em]
 			& BCES Bisector 	     & $8.14 \pm 0.07$ & $3.51 \pm 0.28$ & $0.50$ & $-$ & $0.61$ \\
 			& {\bf mFITEXY Bisector}     & $\boldsymbol{8.15 \pm 0.06}$ & $\boldsymbol{3.37 \pm 0.15}$ & $\boldsymbol{0.50}$ & $-$ & $\boldsymbol{0.59}$ \\
 			& {\tt linmix\_err} Bisector & $8.15 \pm 0.07$ & $3.32 \pm 0.18$ & $0.50$ & $-$ & $0.58$ \\ [0.5em]

 Early-type (E+S0) (43)	& BCES $(Y|X)$  	     & $8.54 \pm 0.10$ & $4.07 \pm 0.87$ & $0.64$ & $-$ & $0.65$ \\
 			& mFITEXY $(Y|X)$	     & $8.58 \pm 0.07$ & $3.32 \pm 0.34$ & $0.64$ & $0.24^{+0.10}_{-0.07}$ & $0.45$ \\
 			& {\tt linmix\_err} $(Y|X)$  & $8.57 \pm 0.08$ & $3.12 \pm 0.43$ & $0.64$ & $0.32 \pm 0.08$ & $0.53$ \\ [0.5em]
 			& BCES $(X|Y)$  	     & $8.54 \pm 0.09$ & $3.95 \pm 0.55$ & $0.64$ & $-$ & $0.63$ \\
 			& mFITEXY $(X|Y)$	     & $8.59 \pm 0.08$ & $3.88 \pm 0.43$ & $0.64$ & $0.26^{+0.11}_{-0.08}$ & $0.62$ \\
 			& {\tt linmix\_err} $(X|Y)$  & $8.59 \pm 0.09$ & $3.82 \pm 0.50$ & $0.64$ & $0.35 \pm 0.10$ & $0.61$ \\ [0.5em]
 			& BCES Bisector 	     & $8.54 \pm 0.10$ & $4.01 \pm 0.63$ & $0.64$ & $-$ & $0.64$ \\
 			& mFITEXY Bisector	     & $8.59 \pm 0.07$ & $3.58 \pm 0.27$ & $0.64$ & $-$ & $0.58$ \\
 			& {\tt linmix\_err} Bisector & $8.58 \pm 0.08$ & $3.44 \pm 0.33$ & $0.64$ & $-$ & $0.56$ \\ [0.5em]

 Spiral (Sp) (17)	& BCES $(Y|X)$  	     & $7.18 \pm 0.28$ & $6.78 \pm 6.62$ & $0.18$ & $-$ & $1.23$ \\
 			& mFITEXY $(Y|X)$	     & $7.24 \pm 0.13$ & $4.48 \pm 0.90$ & $0.18$ & $0.13^{+0.42}_{-0.13}$ & $0.52$ \\
 			& {\tt linmix\_err} $(Y|X)$  & $7.22 \pm 0.16$ & $3.57 \pm 1.36$ & $0.18$ & $0.39 \pm 0.19$ & $0.70$ \\ [0.5em]
 			& BCES $(X|Y)$  	     & $7.18 \pm 0.23$ & $5.48 \pm 1.93$ & $0.18$ & $-$ & $0.99$ \\
 			& mFITEXY $(X|Y)$	     & $7.24 \pm 0.14$ & $4.62 \pm 0.96$ & $0.18$ & $0.13^{+0.43}_{-0.13}$ & $0.85$ \\
 			& {\tt linmix\_err} $(X|Y)$  & $7.21 \pm 0.21$ & $4.86 \pm 1.64$ & $0.18$ & $0.45 \pm 0.31$ & $0.89$ \\ [0.5em]
 			& BCES Bisector 	     & $7.18 \pm 0.25$ & $6.06 \pm 3.66$ & $0.18$ & $-$ & $1.10$ \\
 			& mFITEXY Bisector	     & $7.24 \pm 0.14$ & $4.55 \pm 0.66$ & $0.18$ & $-$ & $0.84$ \\
 			& {\tt linmix\_err} Bisector & $7.22 \pm 0.19$ & $4.12 \pm 1.07$ & $0.18$ & $-$ & $0.77$ \\ [0.5em]

\tableline 
\tableline
\end{tabular}
\label{tab:lregMn} 
\tablecomments{For each subsample, we indicate $\langle \log n_{\rm sph} \rangle$, its average value of spheroid S\'ersic index. 
In the last two columns, we report $\epsilon$, the intrinsic scatter, and $\Delta$, the total rms scatter in the $M_{\rm BH}$ direction. 
The lenticular galaxies NGC 0524 and NGC 3998 were excluded from the linear regression analysis (see Section \ref{sec:anal}). 
Both the early- and late-type subsamples do not contain the two galaxies classified as S0/Sp and the two galaxies classified as mergers (45+17=66-2-2).  }
\end{table*}

  \begin{table*}
  \centering
  \caption{Intrinsic scatter $\epsilon$ of the $M_{\rm BH} - n_{\rm sph}$ and $M_{\rm BH} - L_{\rm sph}$ relations.}
  \begin{tabular}{llcc}
  \tableline
  \tableline
  {\bf Subsample} & {\bf Regression} & $\boldsymbol \epsilon$ {\bf for} $\boldsymbol M_{\rm BH} - n_{\rm sph}$ & 
  $\boldsymbol \epsilon$ {\bf for} $\boldsymbol M_{\rm BH} - L_{\rm sph}$ \\ 
  \tableline 
  \\
  All		  & mFITEXY $(Y|X)$	      & $0.22^{+0.10}_{-0.07}$ & $0.49^{+0.06}_{-0.05}$  \\	     
 		  & {\tt linmix\_err} $(Y|X)$ & $0.29 \pm 0.07$ & $0.51 \pm 0.06$                \\	     
 		  & mFITEXY $(X|Y)$	      & $0.23^{+0.10}_{-0.07}$ & $0.58^{+0.07}_{-0.06}$  \\	     
 		  & {\tt linmix\_err} $(X|Y)$ & $0.30 \pm 0.07$ & $0.60 \pm 0.09$                \\ [0.5em]  
  Early-type	  & mFITEXY $(Y|X)$	      & $0.24^{+0.10}_{-0.07}$ & $0.40^{+0.06}_{-0.05}$  \\	     
 		  & {\tt linmix\_err} $(Y|X)$ & $0.32 \pm 0.08$ & $0.41 \pm 0.06$                \\	     
 		  & mFITEXY $(X|Y)$	      & $0.26^{+0.11}_{-0.08}$ & $0.49^{+0.08}_{-0.06}$  \\	     
 		  & {\tt linmix\_err} $(X|Y)$ & $0.35 \pm 0.10$ & $0.51 \pm 0.10$                \\ [0.5em]  
  Late-type	  & mFITEXY $(Y|X)$	      & $0.13^{+0.42}_{-0.13}$ & $0.55^{+0.15}_{-0.10}$  \\	     
 		  & {\tt linmix\_err} $(Y|X)$ & $0.39 \pm 0.19$ & $0.63 \pm 0.16$                \\	     
 		  & mFITEXY $(X|Y)$	      & $0.13^{+0.43}_{-0.13}$ & $1.09^{+0.41}_{-0.24}$  \\	     
 		  & {\tt linmix\_err} $(X|Y)$ & $0.45 \pm 0.31$ & $1.31 \pm 0.97$                \\ [0.5em]  
  
  \tableline 
  \tableline
  \end{tabular}
  \label{tab:intsc} 
  \end{table*}



\acknowledgments
GS warmly thanks Chris Blake, Alister Graham and Carlo Nipoti for useful discussion. 
We also thank the anonymous referee for useful comments and suggestions. 
This research was supported by Australian Research Council funding through grants
DP110103509 and FT110100263.
This work is based on observations made with the IRAC instrument \citep{fazio2004IRAC} 
on-board the Spitzer Space Telescope, 
which is operated by the Jet Propulsion Laboratory, 
California Institute of Technology under a contract with NASA.
This research has made use of the GOLDMine database \citep{goldmine} and the NASA/IPAC Extragalactic Database (NED) 
which is operated by the Jet Propulsion Laboratory, California Institute of Technology, 
under contract with the National Aeronautics and Space Administration. 
We acknowledge the usage of the HyperLeda database (\url{http://leda.univ-lyon1.fr}).
The BCES routine \citep{akritasbershady1996} was run via the python module 
written by Rodrigo Nemmen \citep{nemmen2012}, which is available at \url{https://github.com/rsnemmen/BCES}.
The modified FITEXY linear regressions were performed using the IDL routine MPFITEXY \citep{williams2010}, 
which is available at \url{http://purl.org/mike/mpfitexy}. 
The MPFITEXY routine depends on the MPFIT package \citep{markwardt2009}.

\bibliography{/Users/gsavorgnan/galaxy_vivisection/papers/SMBHbibliography}

\begin{thebibliography}{}
\expandafter\ifx\csname natexlab\endcsname\relax\def\natexlab#1{#1}\fi

\bibitem[{{Akaike}(1974)}]{akaike1974}
{Akaike}, H. 1974, Automatic Control, IEEE Transactions, 19, 716

\bibitem[{{Akritas} \& {Bershady}(1996)}]{akritasbershady1996}
{Akritas}, M.~G., \& {Bershady}, M.~A. 1996, \apj, 470, 706

\bibitem[{{Andredakis} {et~al.}(1995){Andredakis}, {Peletier}, \&
  {Balcells}}]{andredakis1995}
{Andredakis}, Y.~C., {Peletier}, R.~F., \& {Balcells}, M. 1995, \mnras, 275,
  874

\bibitem[{{Arnold} {et~al.}(2014){Arnold}, {Romanowsky}, {Brodie}, {Forbes},
  {Strader}, {Spitler}, {Foster}, {Blom}, {Kartha}, {Pastorello}, {Pota},
  {Usher}, \& {Woodley}}]{arnold2014}
{Arnold}, J.~A., {Romanowsky}, A.~J., {Brodie}, J.~P., {et~al.} 2014, \apj,
  791, 80

\bibitem[{{Beifiori} {et~al.}(2012){Beifiori}, {Courteau}, {Corsini}, \&
  {Zhu}}]{beifiori2012}
{Beifiori}, A., {Courteau}, S., {Corsini}, E.~M., \& {Zhu}, Y. 2012, \mnras,
  419, 2497

\bibitem[{{Caon} {et~al.}(1993){Caon}, {Capaccioli}, \& {D'Onofrio}}]{caon1993}
{Caon}, N., {Capaccioli}, M., \& {D'Onofrio}, M. 1993, \mnras, 265, 1013

\bibitem[{{Cen}(2014)}]{cen2014}
{Cen}, R. 2014, \apjl, 790, L24

\bibitem[{{Emsellem} {et~al.}(2011){Emsellem}, {Cappellari}, {Krajnovi{\'c}},
  {Alatalo}, {Blitz}, {Bois}, {Bournaud}, {Bureau}, {Davies}, {Davis}, {de
  Zeeuw}, {Khochfar}, {Kuntschner}, {Lablanche}, {McDermid}, {Morganti},
  {Naab}, {Oosterloo}, {Sarzi}, {Scott}, {Serra}, {van de Ven}, {Weijmans}, \&
  {Young}}]{atlas3dIII}
{Emsellem}, E., {Cappellari}, M., {Krajnovi{\'c}}, D., {et~al.} 2011, \mnras,
  414, 888

\bibitem[{{Fazio} {et~al.}(2004){Fazio}, {Hora}, {Allen}, {Ashby}, {Barmby},
  {Deutsch}, {Huang}, {Kleiner}, {Marengo}, {Megeath}, {Melnick}, {Pahre},
  {Patten}, {Polizotti}, {Smith}, {Taylor}, {Wang}, {Willner}, {Hoffmann},
  {Pipher}, {Forrest}, {McMurty}, {McCreight}, {McKelvey}, {McMurray}, {Koch},
  {Moseley}, {Arendt}, {Mentzell}, {Marx}, {Losch}, {Mayman}, {Eichhorn},
  {Krebs}, {Jhabvala}, {Gezari}, {Fixsen}, {Flores}, {Shakoorzadeh}, {Jungo},
  {Hakun}, {Workman}, {Karpati}, {Kichak}, {Whitley}, {Mann}, {Tollestrup},
  {Eisenhardt}, {Stern}, {Gorjian}, {Bhattacharya}, {Carey}, {Nelson},
  {Glaccum}, {Lacy}, {Lowrance}, {Laine}, {Reach}, {Stauffer}, {Surace},
  {Wilson}, {Wright}, {Hoffman}, {Domingo}, \& {Cohen}}]{fazio2004IRAC}
{Fazio}, G.~G., {Hora}, J.~L., {Allen}, L.~E., {et~al.} 2004, \apjs, 154, 10

\bibitem[{{Ferrarese} \& {Merritt}(2000)}]{ferraresemerritt2000}
{Ferrarese}, L., \& {Merritt}, D. 2000, \apjl, 539, L9

\bibitem[{{Gavazzi} {et~al.}(2003){Gavazzi}, {Boselli}, {Donati}, {Franzetti},
  \& {Scodeggio}}]{goldmine}
{Gavazzi}, G., {Boselli}, A., {Donati}, A., {Franzetti}, P., \& {Scodeggio}, M.
  2003, \aap, 400, 451

\bibitem[{{Gebhardt} {et~al.}(2000){Gebhardt}, {Bender}, {Bower}, {Dressler},
  {Faber}, {Filippenko}, {Green}, {Grillmair}, {Ho}, {Kormendy}, {Lauer},
  {Magorrian}, {Pinkney}, {Richstone}, \& {Tremaine}}]{gebhardt2000}
{Gebhardt}, K., {Bender}, R., {Bower}, G., {et~al.} 2000, \apjl, 539, L13

\bibitem[{{Graham}(2001)}]{graham2001bulges}
{Graham}, A.~W. 2001, \aj, 121, 820

\bibitem[{{Graham} \& {Driver}(2005)}]{grahamdriver2005}
{Graham}, A.~W., \& {Driver}, S.~P. 2005, \pasa, 22, 118

\bibitem[{{Graham} \& {Driver}(2007)}]{grahamdriver2007}
---. 2007, \apj, 655, 77

\bibitem[{{Graham} {et~al.}(2003){Graham}, {Erwin}, {Trujillo}, \& {Asensio
  Ramos}}]{graham2003coresersicmodel}
{Graham}, A.~W., {Erwin}, P., {Trujillo}, I., \& {Asensio Ramos}, A. 2003, \aj,
  125, 2951

\bibitem[{{Graham} \& {Guzm{\'a}n}(2003)}]{grahamguzman2003}
{Graham}, A.~W., \& {Guzm{\'a}n}, R. 2003, \aj, 125, 2936

\bibitem[{{Graham} \& {Scott}(2013)}]{grahamscott2013}
{Graham}, A.~W., \& {Scott}, N. 2013, \apj, 764, 151

\bibitem[{{Graham} \& {Worley}(2008)}]{grahamworley2008}
{Graham}, A.~W., \& {Worley}, C.~C. 2008, \mnras, 388, 1708

\bibitem[{{H{\"a}ring} \& {Rix}(2004)}]{haringrix2004}
{H{\"a}ring}, N., \& {Rix}, H.-W. 2004, \apjl, 604, L89

\bibitem[{{Iodice} {et~al.}(1997){Iodice}, {D'Onofrio}, \&
  {Capaccioli}}]{iodice1997}
{Iodice}, E., {D'Onofrio}, M., \& {Capaccioli}, M. 1997, in Astronomical
  Society of the Pacific Conference Series, Vol. 116, The Nature of Elliptical
  Galaxies; 2nd Stromlo Symposium, ed. M.~{Arnaboldi}, G.~S. {Da Costa}, \&
  P.~{Saha}, 84

\bibitem[{{Iodice} {et~al.}(1999){Iodice}, {D'Onofrio}, \&
  {Capaccioli}}]{iodice1999}
{Iodice}, E., {D'Onofrio}, M., \& {Capaccioli}, M. 1999, in Astronomical
  Society of the Pacific Conference Series, Vol. 176, Observational Cosmology:
  The Development of Galaxy Systems, ed. G.~{Giuricin}, M.~{Mezzetti}, \&
  P.~{Salucci}, 402

\bibitem[{{Jerjen} {et~al.}(2000){Jerjen}, {Binggeli}, \&
  {Freeman}}]{jerjen2000}
{Jerjen}, H., {Binggeli}, B., \& {Freeman}, K.~C. 2000, \aj, 119, 593

\bibitem[{{Kelly}(2007)}]{linmixerr}
{Kelly}, B.~C. 2007, \apj, 665, 1489

\bibitem[{{Khosroshahi} {et~al.}(2000){Khosroshahi}, {Wadadekar}, \&
  {Kembhavi}}]{khosroshahi2000}
{Khosroshahi}, H.~G., {Wadadekar}, Y., \& {Kembhavi}, A. 2000, \apj, 533, 162

\bibitem[{{Kormendy} \& {Richstone}(1995)}]{kormendyrichstone1995}
{Kormendy}, J., \& {Richstone}, D. 1995, \araa, 33, 581

\bibitem[{{Magorrian} {et~al.}(1998){Magorrian}, {Tremaine}, {Richstone},
  {Bender}, {Bower}, {Dressler}, {Faber}, {Gebhardt}, {Green}, {Grillmair},
  {Kormendy}, \& {Lauer}}]{magorrian1998}
{Magorrian}, J., {Tremaine}, S., {Richstone}, D., {et~al.} 1998, \aj, 115, 2285

\bibitem[{{Marconi} \& {Hunt}(2003)}]{marconihunt2003}
{Marconi}, A., \& {Hunt}, L.~K. 2003, \apjl, 589, L21

\bibitem[{{Markwardt}(2009)}]{markwardt2009}
{Markwardt}, C.~B. 2009, in Astronomical Society of the Pacific Conference
  Series, Vol. 411, Astronomical Data Analysis Software and Systems XVIII, ed.
  D.~A. {Bohlender}, D.~{Durand}, \& P.~{Dowler}, 251

\bibitem[{{Nemmen} {et~al.}(2012){Nemmen}, {Georganopoulos}, {Guiriec},
  {Meyer}, {Gehrels}, \& {Sambruna}}]{nemmen2012}
{Nemmen}, R.~S., {Georganopoulos}, M., {Guiriec}, S., {et~al.} 2012, Science,
  338, 1445

\bibitem[{{Nipoti}(2015)}]{nipoti2015}
{Nipoti}, C. 2015, \apjl, 805, L16

\bibitem[{{Novak} {et~al.}(2006){Novak}, {Faber}, \& {Dekel}}]{novak2006}
{Novak}, G.~S., {Faber}, S.~M., \& {Dekel}, A. 2006, \apj, 637, 96

\bibitem[{{Park} {et~al.}(2012){Park}, {Kelly}, {Woo}, \& {Treu}}]{park2012}
{Park}, D., {Kelly}, B.~C., {Woo}, J.-H., \& {Treu}, T. 2012, \apjs, 203, 6

\bibitem[{{Press} {et~al.}(1992){Press}, {Teukolsky}, {Vetterling}, \&
  {Flannery}}]{press1992}
{Press}, W.~H., {Teukolsky}, S.~A., {Vetterling}, W.~T., \& {Flannery}, B.~P.
  1992, {Numerical recipes in FORTRAN. The art of scientific computing}

\bibitem[{{Rusli} {et~al.}(2013){Rusli}, {Erwin}, {Saglia}, {Thomas},
  {Fabricius}, {Bender}, \& {Nowak}}]{rusli2013}
{Rusli}, S.~P., {Erwin}, P., {Saglia}, R.~P., {et~al.} 2013, \aj, 146, 160

\bibitem[{{Sani} {et~al.}(2011){Sani}, {Marconi}, {Hunt}, \&
  {Risaliti}}]{sani2011}
{Sani}, E., {Marconi}, A., {Hunt}, L.~K., \& {Risaliti}, G. 2011, \mnras, 413,
  1479

\bibitem[{{Savorgnan} {et~al.}(2013){Savorgnan}, {Graham}, {Marconi}, {Sani},
  {Hunt}, {Vika}, \& {Driver}}]{savorgnan2013}
{Savorgnan}, G., {Graham}, A.~W., {Marconi}, A., {et~al.} 2013, \mnras, 434,
  387

\bibitem[{{Savorgnan} \& {Graham}(2015{\natexlab{a}})}]{ellicular}
{Savorgnan}, G.~A.~D., \& {Graham}, A.~W. 2015{\natexlab{a}}, ArXiv e-prints,
  arXiv:1511.05654

\bibitem[{{Savorgnan} \& {Graham}(2015{\natexlab{b}})}]{paperI}
---. 2015{\natexlab{b}}, ArXiv e-prints, arXiv:1511.07446

\bibitem[{{Savorgnan} {et~al.}(2015){Savorgnan}, {Graham}, {Marconi}, \&
  {Sani}}]{paperII}
{Savorgnan}, G.~A.~D., {Graham}, A.~W., {Marconi}, A., \& {Sani}, E. 2015,
  ArXiv e-prints, arXiv:1511.07437

\bibitem[{{Scott} {et~al.}(2014){Scott}, {Davies}, {Houghton}, {Cappellari},
  {Graham}, \& {Pimbblet}}]{scott2014}
{Scott}, N., {Davies}, R.~L., {Houghton}, R.~C.~W., {et~al.} 2014, \mnras, 441,
  274

\bibitem[{{Seigar} \& {James}(1998)}]{seigar1998}
{Seigar}, M.~S., \& {James}, P.~A. 1998, \mnras, 299, 672

\bibitem[{{S{\'e}rsic}(1963)}]{sersic1963}
{S{\'e}rsic}, J.~L. 1963, Boletin de la Asociacion Argentina de Astronomia La
  Plata Argentina, 6, 41

\bibitem[{{S{\'e}rsic}(1968)}]{sersic1968}
---. 1968, {Atlas de galaxias australes}

\bibitem[{{Sheth} {et~al.}(2010){Sheth}, {Regan}, {Hinz}, {Gil de Paz},
  {Men{\'e}ndez-Delmestre}, {Mu{\~n}oz-Mateos}, {Seibert}, {Kim},
  {Laurikainen}, {Salo}, {Gadotti}, {Laine}, {Mizusawa}, {Armus},
  {Athanassoula}, {Bosma}, {Buta}, {Capak}, {Jarrett}, {Elmegreen},
  {Elmegreen}, {Knapen}, {Koda}, {Helou}, {Ho}, {Madore}, {Masters},
  {Mobasher}, {Ogle}, {Peng}, {Schinnerer}, {Surace}, {Zaritsky},
  {Comer{\'o}n}, {de Swardt}, {Meidt}, {Kasliwal}, \& {Aravena}}]{sheth2010}
{Sheth}, K., {Regan}, M., {Hinz}, J.~L., {et~al.} 2010, \pasp, 122, 1397

\bibitem[{{Tremaine} {et~al.}(2002){Tremaine}, {Gebhardt}, {Bender}, {Bower},
  {Dressler}, {Faber}, {Filippenko}, {Green}, {Grillmair}, {Ho}, {Kormendy},
  {Lauer}, {Magorrian}, {Pinkney}, \& {Richstone}}]{tremaine2002}
{Tremaine}, S., {Gebhardt}, K., {Bender}, R., {et~al.} 2002, \apj, 574, 740

\bibitem[{{Trujillo} {et~al.}(2004){Trujillo}, {Erwin}, {Asensio Ramos}, \&
  {Graham}}]{trujillo2004coresersicmodel}
{Trujillo}, I., {Erwin}, P., {Asensio Ramos}, A., \& {Graham}, A.~W. 2004, \aj,
  127, 1917

\bibitem[{{Vika} {et~al.}(2012){Vika}, {Driver}, {Cameron}, {Kelvin}, \&
  {Robotham}}]{vika2012}
{Vika}, M., {Driver}, S.~P., {Cameron}, E., {Kelvin}, L., \& {Robotham}, A.
  2012, \mnras, 419, 2264

\bibitem[{{Williams} {et~al.}(2010){Williams}, {Bureau}, \&
  {Cappellari}}]{williams2010}
{Williams}, M.~J., {Bureau}, M., \& {Cappellari}, M. 2010, \mnras, 409, 1330

\bibitem[{{Young} \& {Currie}(1994)}]{youngcurrie1994}
{Young}, C.~K., \& {Currie}, M.~J. 1994, \mnras, 268, L11

\end{thebibliography}

\clearpage

\end{document}